\documentclass[10pt]{article}
\usepackage{makeidx}
\usepackage{multirow}
\usepackage{multicol}
\usepackage[dvipsnames,svgnames,table]{xcolor}
\usepackage{graphicx}
\usepackage{ulem}
\usepackage{amsmath}
\usepackage{amssymb}
\author{Noah}
\title{}
\usepackage[paperwidth=612pt,paperheight=792pt,top=72pt,right=72pt,bottom=72pt,left=72pt]{geometry}

\makeatletter
	{\par\setlength{\parindent}{#3}
	\setlength{\leftmargin}{#1}       \setlength{\rightmargin}{#2}%
	\advance\linewidth -\leftmargin       \advance\linewidth -\rightmargin%
	\advance\@totalleftmargin\leftmargin  \@setpar{{\@@par}}%
	\parshape 1\@totalleftmargin \linewidth\ignorespaces}{\par}%
\makeatother

\begin{document}

\title{Two Steps to Obfuscation}
\author{Noah E. Friedkin\textsuperscript{1} and Eugene C. Johnsen\textsuperscript{2}} 
\maketitle

\noindent \textit{ \textsuperscript{1} Center for Control, Dynamical Systems and Computation, College of Engineering, and  Department of Sociology, College of Letters and Science, \textsuperscript{2} Department of Mathematics, University of California, Santa Barbara  }.

\begin{abstract} 
\noindent This note addresses the historical antecedents of the 1998 PageRank measure of centrality. An identity relation links it to 1990-1991 models of Friedkin and Johnsen.   
\end{abstract}

\section{Introduction}

Friedkin and Johnsen (1990) presented a model of a multi-agent network in which the total influences of the agents are related to the number and length of the walks in the network \cite{FJ1990} as follows  

\begin{align}
{\bf{V}} & = ({\bf{I}} + \alpha{\bf{W}} + {{\alpha}^2}{\bf{W}}^{2} + {{\alpha}^3}{\bf{W}}^{3} + ... )(1- \alpha) \notag \\
& = ({\bf{I}} - \alpha{\bf{W}})^{-1} (1- \alpha)
\label{eq:powerseries}
\end{align}

\noindent where ${\bf{W}}_{n \times n}$ is row-stochastic and  $0<\alpha<1$ is scalar. This formulation was novel, and seminal to subsequent work in which the homogeneity of the $1-\alpha$ factor was relaxed \cite{Friedkin1998,FJ1999,FJ2011}. The $v_{ij}$ of ${\bf{V}}=[v_{ij}]$ corresponds to the relative net influence of agent $j$ on agent $i$. 

Friedkin (1991) developed the employment of ${\bf{V}}$ as a measure of structural centrality 

\begin{align} 
{\bf{c}} & = \frac{1}{n}{{\bf{V}}^{T}}{\bf{e}} , \;\; 
{{\bf{c}}^{T}} {\bf{e}} = 1,
\label{eq:TEC1}  \\
& = \bigg( \frac{1- \alpha}{n} \bigg) \bigg( {\bf{I}} - \alpha{\bf{W^{T}}}\bigg)^{-1} {\bf{e}}
\label{eq:TEC}
\end{align}

\noindent where here, and henceforth, {\bf{e}} is a vector of ones and each element of ${\bf{c}}$ is ``the average total effect centrality of an actor" \cite[pp.1485 -1487]{Friedkin1991}.  The average may be based on the $n$ values of each column, or $n-1$ values when the main diagonal values of ${\bf{V}}$ are excluded. The latter concentrates the measure on the total effects of an agent $i$ on \textit{other} agents. 

If the vector of averages in equation \ref{eq:TEC1} are expressed as follows

\begin{equation} 
{\bf{c}}  = \bigg( \frac{1- \alpha}{n} \bigg){\bf{e}}  + \alpha{\bf{W}}^{T} {\bf{c}}, 
\label{eq:PageRank}
\end{equation}

\noindent then equation \ref{eq:TEC} are their solutions. If no averages are taken, then the model simply presents the sums of the columns of ${\bf{V}}$

\begin{align} 
{\bf{c}} & = {{\bf{V}}^{T}}{\bf{e}} ,  \notag \\
& = \big( {1- \alpha} \big) \bigg( {\bf{I}} - \alpha{\bf{W^{T}}}\bigg)^{-1} {\bf{e}}, 
\label{eq:Vsums}
\end{align}
\noindent whence

\begin{equation}
{\bf{c}} = ({1- \alpha}){\bf{e}}  + \alpha{\bf{W}}^{T} {\bf{c}}.
\label{eq:PRVsums}
\end{equation}

\noindent We will now show why this odd form of the model (equation \ref{eq:PRVsums}) is of interest.

\section{Two steps to obfuscation}

Consider an application of the model to a webgraph composed of nodes that are the pages of the webgraph and edges that are its hyperlinks. Let  ${\bf{A}}=[a_{ij}]$ be the adjacency matrix of the webgraph, where ${a_{ij}} =1$ if page $i$ has a directed link to page $j$ and 0 otherwise. Let ${\bf{W}}=[w_{ij}]$ be the normalized adjacency matrix, 

\begin{equation}
{w_{ij}} = \frac{a_{ij}}{{\sum_{k=1}^n}{a_{ik}}} 
= \frac{a_{ij}}{od(i)}, \;\; {od(i)}>0
\end{equation}   

\noindent for all $i$ and $j$. Equation \ref{eq:PRVsums} may now be expressed as follows

\begin{equation}
c_{i} = ({1- \alpha})  + \alpha \sum_{j \in S}^{n}{\frac{c_{j}}{od(j)}}, \;\;  {od(j)} >0,
\end{equation} 

\noindent for all $i$, where $S$ is the set of edges for which $j$ has a direct link to $i$.

Step 1. Now alter the notation. Let $PR(j) \equiv {c_{j}}, \; j=1,..,n$ and let $d \equiv \alpha$. Those changes of notation present

\begin{equation}
PR(i) = (1- d)  + d \sum_{j \in S}^{n}{\frac{PR(j)}{od(j)}}, \;\; {od(j)} >0
\end{equation} 

Step 2. Now alter the remaining notation. Let $A$ be $i$ and let  $T1,...,Tn$ be the $j=1,..,n$ pages that point to it. Let $C(A)$ be the number of links going out of page $A$. Those changes of notation present

\begin{equation}
PR(A) = (1- d)  + d \bigg[ \frac{PR(T1)}{C(T1)}+...+ \frac{PR(Tn)}{C(Tn)} \bigg] 
\label{eq:BPsum} 
\end{equation} 
 
\noindent which is exactly the description of the PageRank calculation that Page and Brin (1998) presented as the foundation of Google \cite{Brin1998}. It is equivalent to equation \ref{eq:PRVsums}, and equation  \ref{eq:Vsums} is its solution, i.e., the unnormalized measure of centrality. However, the Page and Brin presentation of it generated some confusion and it was subsequently modified to the normalized measure

\begin{equation}
PR(A) = \bigg( \frac{1-d}{n} \bigg) + d \bigg[ \frac{PR({T1})}{C(T1)}+...+ \frac{PR(Tn)}{C(Tn)} \bigg]  
\end{equation} 

\noindent This equation is equivalent to equation \ref{eq:PageRank}, and equation \ref{eq:TEC} is its solution. In either case, equation \ref{eq:powerseries} provides the foundation of an algorithmic approximation of ${\bf{V}}$ when  inverse computations are not feasible.

These two sets of notation have completely obscured the  equivalence of the PageRank calculations and the constructs of Friedkin and Johnsen (1990) and Friedkin (1991). Mathematica now presents a PageRank centrality solution for the adjacency matrices of digraphs that, with $\alpha = 0.85$, returns centrality scores that are identical to those of equation \ref{eq:TEC}.

\section{Discussion}

The PageRank formula was published over fifteen years ago in a venue outside of sociology. This is ``ancient'' history. But it is a history that remains relevant today whenever the solution of the formula is employed as a measure of centrality. Reinvention of the wheel in the field of social networks is not an unfamiliar event now that investigators from the natural and engineering sciences have become more interested in social networks. Notation differences, and the difficulty of monitoring publications appearing in journals outside one's own discipline, have obscured the correspondence of the 1998 PageRank measure and the 1991 measure proposed by Friedkin.

Brin and Page presented the formula as an intuitive hop to a novel  eigenvector-like measure. In contrast, Friedkin's 1991 measure was a development of Friedkin and Johnsen's 1990 model. The 1990 construct ${\bf{V}}$ was an analytically derived corollary of their specification of a proposed convex combination mechanism of influence among agents joined in a multi-agent network. The 1990 model, in turn, was developed as a generalization of the seminal work of French \cite{French1956}. An eigenvector approach to centrality is natural and appealing. The relaxed eigenvector-like formula of equations \ref{eq:PageRank} and \ref{eq:PRVsums}, and their PageRank equivalents, do not appear in our work as an intuitive hop.

\bibliography{ArticleReferences}

\end{document}